\documentclass[12pt]{article}

\usepackage{times}
\usepackage{epsfig}
\usepackage{amsmath}
\usepackage{color}

\newenvironment{sciabstract}{%
\begin{quote} \bf}
{\end{quote}}

\title{Onset of natural selection in 
auto-catalytic heteropolymers} 

\author
{Alexei V. Tkachenko,$^{1\ast}$ and Sergei Maslov$^{2\ast}$
\\
\normalsize{$^{1\ast}$Center for Functional Nanomaterials, 
Brookhaven National Laboratory, Upton NY 11973,}\\
\normalsize{$^{2\ast}$Department of Bioengineering and Carl R. Woese Institute for Genomic Biology,}\\
\normalsize{University of Illinois at Urbana-Champaign, Urbana, IL 61801}\\
\normalsize{$^\ast$To whom correspondence should be addressed;}\\ 
\normalsize{
E-mails:  oleksiyt@bnl.gov (AVT) and maslov@illinois.edu (SM)}
}

\date{}

\begin{document} 


\baselineskip24pt


\maketitle 
\begin{sciabstract}
Reduction of information entropy along with 
ever-increasing complexity are among the key signatures of living matter. 
Understanding the onset of such behavior 
in early prebiotic world is essential for solving the problem of  origins of life.
To elucidate this transition, we study a theoretical model 
of information-storing heteropolymers capable of 
template-assisted ligation and subjected to cyclic non-equilibrium 
driving forces. We discover that this simple physical 
system undergoes a spontaneous reduction of the information entropy 
due to the competition of chains for constituent monomers. This 
natural-selection-like process ultimately results in the survival 
of a limited subset of polymer sequences. 
Importantly, the number of surviving sequences remains exponentially large, thus  
opening up the possibility of further increase in complexity 
due to  Darwinian evolution. We also propose potential experimental 
implementations of our model using either biopolymers or 
artificial nano-structures.
%
%
\end{sciabstract}


\section*{Introduction}
The second law of thermodynamics states that the entropy of a closed system 
increases with time. Life represents a remarkable example of the opposite 
trend taking place in an open, non-equilibrium system \cite{lovelock}. 
Indeed, both information 
and thermodynamic entropies decrease in the course of Darwinian evolution 
reflecting ever-increasing complexity of living organisms and their communities
\cite{disclaimer}. 
Interestingly, both the second law and the concept of entropy were introduced by 
Clausius in 1850s, roughly at the same time as Darwin developed and published 
his seminal work. 
A century later the connection between life and entropy was highlighted in the classical 
work of Schr\"{o}dinger titled "What is life?" \cite{schroedinger}. According to him, living systems 
are characterized by their ability to "feed on" and store the negative entropy (which he referred to 
as "negentropy") \cite{disclaimer}. 
In the same work, he effectively predicted the existence 
of information-storing molecules such as DNA. Soon after, Brillouin established 
\cite{brillouin}
the connection between the thermodynamic negentropy and its information cousin 
defined by Shannon \cite{shannon}.   

The emergence of life from non-living matter is one of the greatest 
mysteries of fundamental science 
In addition, the search for artificial self-replicating 
nano- and micro-scale systems is an exciting field with potential 
engineering applications \cite{Chaikin,Brenner2014,Brenner2017,Delacruz2015}. The central challenge in both of 
these fields is to come up with 
a simple physically-realizable self-replicating system obeying the laws of thermodynamics,
yet ultimately capable of Darwinian evolution when exposed 
to non-equilibrium driving forces.

Chemical networks of molecules engaged in the mutual catalysis have long been considered a plausible form of prebiotic world \cite{Eigen,Dyson,Kauffman,Jain}. 
  Furthermore, a set of mutually-catalyzing 
RNA-based enzymes (ribozymes) is one of the best known examples of experimentally realized  autonomous self-replication. This is viewed as a major 
evidence supporting the RNA-world hypothesis (see e.g. Refs. \cite{Joyce_Review}-\cite{Orgel}). 
The ribozyme activity requires relatively 
long polymers made of hundreds of nucleotides with 
carefully designed sequences whose spontaneous emergence by pure 
chance is nearly impossible. 
%
%
Thus, to make the first steps towards 
explanation of the origin of life, one needs to come up with 
a much simpler system capable of spontaneous
reduction of the information entropy, that would ultimately 
set the stage for Darwinian evolution e.g. towards functional 
ribozymes and/or autocatalytic metabolic cycles.

\begin{figure}
\includegraphics[angle=0,width=0.95\columnwidth]{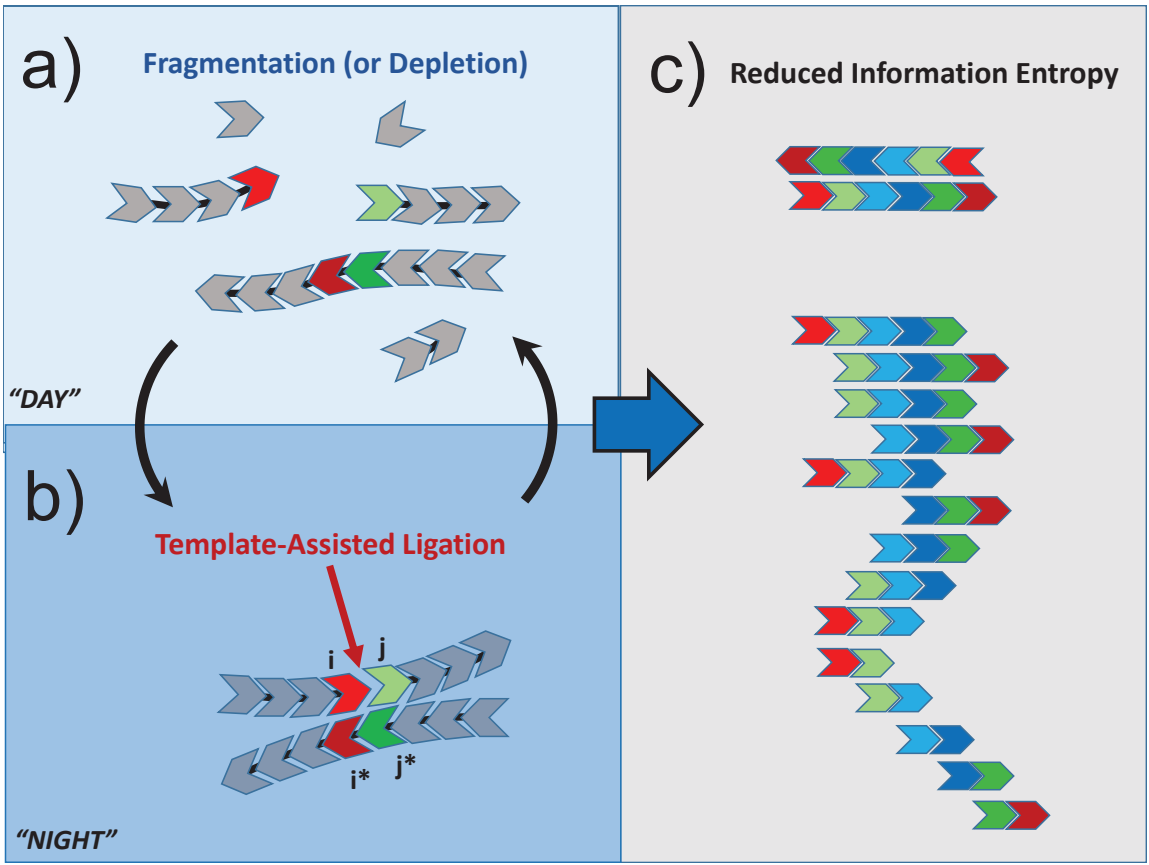}
\caption{
A conceptual representation of our model in which the system is cycled between day, a), and 
night, b), phases. During the night phase polymer chains undergo template-assisted ligation 
joining left and right ends $i$ and $j$ respectively to form a new 2-mer $ij$.
This process is assisted by a complementary 2-mer $j^*i^*$.
The process results in a reduced information entropy of chain sequences dominated 
by a small subset of 2-mers illustrated in c).
\label{fig1}
}
\end{figure}
A promising candidate 
for such mechanism is provided by template-assisted ligation. 
In this process pairs of polymers are brought together 
via hybridization with a complementary template chain 
and eventually ligated to form a longer chain
(see Fig. \ref{fig1}a-b). 
Unlike the 
non-templated reversible step-growth polymerization 
used in Ref. \cite{Braun}, 
this mechanism naturally involves the transmission of an information 
from a template to the newly-ligated chain, thus opening an exciting 
possibility of long-term memory and evolvability.
An early conceptual model involving template-assisted polymerization
was proposed by P. W. Anderson and 
colleagues \cite{Anderson,Anderson_Stein}. 
It  has also been a
subject of several more recent experimental and theoretical 
studies \cite{Szostak_1996,Derr,Hordijk2011}. 
In particular, the model by Hordijk et al. 
\cite{Hordijk2011} makes a connection between 
the classical Kauffman model of autocatalytic sets 
\cite{Kauffman} and polymer systems capable of 
template-assisted ligation (see also \cite{Fellermann2017} for 
further development of that approach). 
Recently, we theoretically established \cite{tkachenko2015}
that a cyclically-driven system of this type is 
capable of producing long, mutually-catalyzing chains starting from a primordial soup 
dominated by monomers. 
%
In the current study we focus on statistics of sequences 
of these chains
and discover that the dynamics of the system naturally 
results in a dramatic reduction of the information entropy 
in the sequence space. 

\section*{Model}

Here we further develop the model introduced in 
Ref. \cite{tkachenko2015}.
It describes the emergence of heteropolymers 
out of the ``primordial soup'' of 
monomers by  virtue of template-assisted 
ligation. 
%
Our system 
is driven out of equilibrium by cyclic changes in 
physical conditions such as temperature, 
salt concentration, pH, etc. (see Fig. \ref{fig1}).

We consider a general case of information-coding heteropolymers 
composed of $Z$ types of monomers 
capable of making $Z/2$  mutually complementary pairs.
%
Polymerization occurs during the "night" phase
of each cycle when  existing heteropolymers may serve as templates 
for ligation of pairs of chains to form longer ones. 
%
When the end groups of two substrate chains  
are positioned next to each other
by the virtue of hybridization with the 
template, a new covalent bond connecting these end groups is
formed at a certain rate (see Fig. \ref{fig1}b). 
During the "day" phase of each cycle 
all hybridized pairs dissociate, and individual chains 
are fully dispersed (see Fig. \ref{fig1}a).

One of the key results of our previous work \cite{tkachenko2015}
is the existence of the 
optimal hybridization overlap length $k_0$ for 
template-substrate binding. 
In this work, for the sake of simplicity 
we assume that 
{\it a single pair} of complementary monomers is sufficient 
to bind a substrate to the template. This can be interpreted as 
if each of $Z$ monomers in the present model is in fact a "word" 
composed of $k_0$ smaller elementary letters, 
such as e.g. RNA or DNA bases. 
Within this interpretation, the number $Z$ of such "composite monomers" 
can be exponentially large: $Z=z^{k_0}$, where $z$ is the number of 
elementary letters ($z=4$ in the case of RNA).
%
%
As in \cite{tkachenko2015} we ignore 
the process of spontaneous, non-templated 
ligation \cite{Derr,Braun}.  
%
%
 


In our model monomer types are labeled 
in such a way that type $i$ is complementary to type  $i^*$.
One of the key concepts in our analysis is that of a 
"2-mer" $ij$ referring to a monomer $i$ immediately followed by the monomer $j$ and 
{\it found anywhere within any heteropolymer}. Note that, similar to 
DNA/RNA complementary strands, polymers in our system are 
assumed to be directional and anti-parallel when hybridized to each other. Therefore, 
a 2-mer $j^*i^*$ formed from monomers $j^*$ and $i^*$ is complementary to the 
2-mer $ij$. 
%
It can serve as a template catalyzing the ligation 
of two substrate chains with monomers $i$ and $j$ located 
at their appropriate ends. 
(see Fig. \ref{fig1}b).

Let $d_{ij}$ denote the overall concentration of 2-mers of type $ij$, 
i.e. the total number of consecutive monomers of types $i$ and $j$ 
found anywhere within any chain, divided by the volume of the system.
We will refer to the $Z \times Z$ matrix formed by all $d_{ij}$ 
as the 2-mer matrix.   
Let  $l_i$ denote the concentration of all chains ending with a 
monomer of type $i$, while 
$r_j$ - the concentration of all chains starting with a monomer of type $j$.
When two ends $i$ and $j$ of such chains meet due to hybridization 
with a complementary template $j^* i^*$, they are 
ligated at a certain rate to form a new 2-mer $ij$. 
We describe this process 
by a three-body mass-action term 
$\lambda_{ij} \cdot l_i(t) \cdot r_j(t) \cdot d_{j^*i^*}(t)$.
Here $\lambda_{ij}$ is the ligation rate averaged over the duration of the 
day-night cycle with the understanding that it happens only 
during the night phase. 
2-mers $ij$ in our system are assumed to spontaneously break up at a rate
$\beta_{ij}$.
Thus we extend the original model 
by introducing an  explicit sequence dependence of 
ligation ($\lambda_{ij}$) and breakage 
($\beta_{ij}$) rates. 
Master equations, describing the slow dynamics in 
our system occurring over multiple day/night cycles, are:
\begin{equation}
\dot{d}_{ij}(t) = \lambda_{ij} l_i(t) r_j(t) d_{j^*i^*}(t) 
- \beta_{ij} d_{ij}(t)  \qquad .
\label{eq_2-mer_dynamics}
\end{equation}
This mass-action description implies that our system stays well below the 
saturation regime during the night phase. In other words, we assume that 
template-substrate hybridization probability is determined by the 
association rate and not by the competition of multiple different substrates 
for the same binding site on a chain. This is realized when the duration of the night phase 
of the cycle is shorter than the typical association time for hybridization. 
Importantly, this regime also ensures that there 
is no template poisoning, i.e. the probability of two complementary 2-mers binding 
each other (and thus loosing their catalytic activity) remains low. 
%
%
%

One could write a similar set of kinetic equations describing 
the dynamics of concentrations of 
``left'' and ``right'' ends of chains, $l_i(t)$ and $r_i(t)$. 
Instead, we use the conservation of overall concentrations 
of monomers of each type to obtain the explicit algebraic expressions for 
$l_i(t)$ and $r_i(t)$ in terms of the 2-mer matrix:
\begin{eqnarray} 
l_i(t)=c_i-\sum_k d_{ik}(t) \quad , \nonumber \\
r_i(t)=c_i-\sum_k d_{ki}(t) \quad .
\label{eq_li} 
\end{eqnarray}
Here $c_i$ is the overall concentration of monomers of type $i$ in the pool, 
both free and bound. At the start only free monomers are present ($d_{ij}=0$), 
and thus the initial conditions are given by $l_i(0)=r_i(0)=c_{i}$. 

%
%
%
Our model allows for an alternative interpretation that 
does not involve breakage of intra-polymer bonds. 
Specifically, as shown in the Supplementary Materials (SM), 
Eqs. (\ref{eq_2-mer_dynamics}-\ref{eq_li}) also describe 
a system subject to uniform dilution at rate $\beta$ and 
the influx of fresh monomers $\phi_i=\beta \cdot c_i$. 
In the light of this interpretation, below we 
focus on the case of all $\beta_{ij}$ 
equal to each other. Without loss 
of generality, they can all  be set to unity: $\beta_{ij}=1$. 
This defines the fundamental timescale in our system as either 
the average lifetime of a single bond or the inverse of the dilution rate.
%

\section*{Results}
\subsection*{Spontaneous entropy reduction} 
In our previous study \cite{tkachenko2015} we worked within random sequence approximation. 
If all monomers have identical total concentrations $c_i=c$, 
this approximation corresponds to all matrix elements $d_{ij}(t)$ being 
equal to each other. For general initial conditions these elements would be proportional to 
$c_{i} \cdot c_{j}$. The key hypothesis proposed but not tested in Ref. \cite{tkachenko2015} 
is that the system dynamics would eventually favor the survival of a subset of the "fittest" 
sequences at the expense of the others, thus breaking the random sequence approximation. 
Here we test this hypothesis by simulating the dynamics of the model given by 
Eqs. (\ref{eq_2-mer_dynamics}-\ref{eq_li}) with $Z=20$. We start with a system characterized by a
weak variation in individual ligation rates $\lambda_{ij}$ and concentrations $c_i$. We choose 
them from a log-normal distribution with their logarithms having standard deviation 0.1 and means 
$0$ and $\log 3$ respectively. Our choice of parameters is motivated by the need to understand the 
the limit of infinitesimally weak variation of rates and concentrations.
For this combination of parameters, the Eqs. \ref{eq_2-mer_dynamics} 
are initially linearly unstable with respect to formation of all 2-mers.
However, no 2-mer would be formed until either it or its complementary 
partner is present in the system at least in some infinitesimal "seed" 
concentration. Once such seed is introduced, the corresponding pair of mutually 
complementary 2-mers $ij$ and $j^*i^*$
would be exponentially amplified. 
In our simulations we used the same small seed concentration of $10^{-4}$ for all $Z^2$ 2-mers

\begin{figure}
\begin{center}
\includegraphics[angle=0,width=0.95\columnwidth]{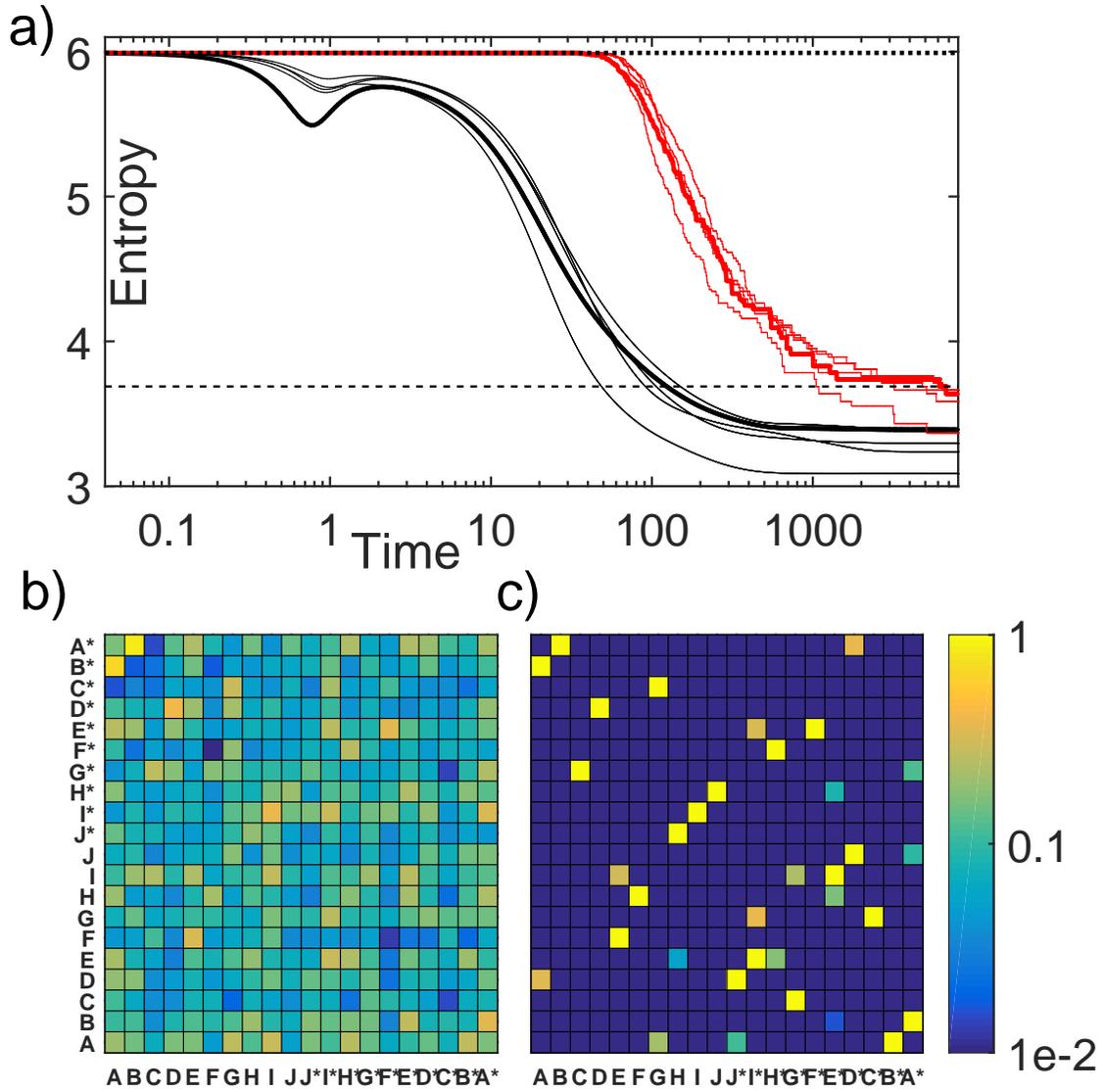}
\end{center}
\caption{a) The information entropy $S$ (black lines) given by Eq. \ref{entropy} and the natural 
logarithm of the number of surviving 2-mers $N$ (red lines) plotted 
vs time in 5 different realizations of our model with logarithms of 
both $\lambda_{ij}$ and $c_i$ normally distributed with 
standard deviation of $0.1$ and means of $\log 1$ and $\log 3$ respectively. 
b) The heatmap visualizing $\log_{10}$ of concentrations of 2-mers at $t=2$ 
(the second maximum of the entropy) in one of these realizations highlighted by 
thick black and red lines in panel a). 
c) The same heatmap in the steady state at $t=8000$ where the entropy is 
saturated at its lowest point.
\label{fig2}
}
\end{figure}

The key parameter we use to quantify the emergent complexity in our system is the 
information entropy of 2-mers 
based on their relative concentrations 
$\tilde{d}_{ij}=d_{ij}/\sum_{kl}d_{kl}$
and defined in the standard Boltzmann-Shannon manner:
\begin{equation}
S(t)=-\sum_{kl}\tilde{d}_{kl}(t) \log \tilde{d}_{kl}(t) \qquad .
\label{entropy}
\end{equation}
Fig. \ref{fig2} shows the time dependence of this entropy
in 5 different realizations of $\lambda_{ij}$ and $c_i$.
The entropy starts at its maximal value $\log(Z^2)$, 
and after a brief dip followed by a rebound, it
steadily {\it declines} as a function of time. 
Such behavior is a remarkable manifestation of the non-equilibrium nature 
of our system, as the entropy changes in the direction opposite to that 
dictated by the second law of thermodynamics. 
To reveal the source of this entropy dynamics, in Figs. \ref{fig2}b and \ref{fig2}c 
we show the 2-mer matrix at two time points during our simulations. 
At $t=2$ all of the 
2-mers have grown up from their seed to substantial concentrations. 
Remarkably, the subsequent dynamics 
leads to a {\it complete extinction} of the majority of 2-mers ultimately 
giving rise to the 2-mer matrix at $t=8000$ shown in Fig. \ref{fig2}c. The time 
dependence of the logarithm of the number of surviving 
2-mers is shown as red lines in Fig. \ref{fig2}a. The ultimate 
number of survivors $36\pm 4$ is just below $2Z=40$ (out of $Z^2=400$) represented by 
the lower horizontal dotted line at $\log 2Z$ in Fig. \ref{fig2}a. 

\subsection*{Competition between 2-mers and the number of survivors}
The observed  behavior can be understood from the analysis of Eqs. 
(\ref{eq_2-mer_dynamics}-\ref{eq_li}). 
%
For a fixed set of concentrations $l_i$ and $r_i$, 
Eqs. \ref{eq_2-mer_dynamics} form a set of linear 
kinetic  equations with respect to 2-mer concentrations $d_{ij}$. 
Furthermore, this set of $Z^2$ equations breaks into independent 
blocks of equations describing the dynamics of mutually 
complementary 2-mers $d_{ij}$ and $d_{j*i*}$. For a small subset 
of self-complementary 2-mers $ii^*$, occupying a diagonal of the 
2-mer matrix, such a block is represented by a single equation. 
In all other cases it involves a pair of equations for $d_{ij}$ 
and $d_{j*i*}$ coupled via a 2x2 matrix:
\begin{equation}
\begin{pmatrix} \dot{d}_{ij} \\ \dot{d}_{j*i*}\end{pmatrix} =
\begin{pmatrix} -1 & \lambda_{ij} l_i r_j\\ \lambda_{j*i*} l_{j*} r_{i*}& -1\end{pmatrix}
\begin{pmatrix} d_{ij} \\ d_{j*i*} \end{pmatrix} 
\label{matrix}
\end{equation}
%
Because the trace of the matrix is always negative, at 
least one of the eigenvalues has to have a negative real part, 
while the real part of the other one could be positive, 
negative, or zero depending on the value of the matrix determinant
$\Delta_{ij}$. 
A negative value of the determinant $\Delta_{ij}<0$ corresponds 
to a positive eigenvalue and hence to the exponential growth of two 
complementary 2-mer concentrations observed at the initial stage. 
As growing 2-mers gradually deplete $l_i$ and $r_j$, $\Delta_{ij}$ 
increases and may eventually turn positive. In this case both 
eigenvalues become negative. This triggers the exponential decay of 
concentrations and ultimate extinction of 
the corresponding pair of 2-mers. 
A small subset of 2-mers survive and reach the steady state. For these survivors, 
the determinant {\it has to become exactly zero}: $\Delta_{ij}=0$. 
These conditions for surviving 2-mers can be rewritten as 
\begin{equation}
\Delta_{ij} \equiv 1-\lambda_{ij}l_ir_j \cdot \lambda_{j*i*}r_{i*}l_{j*}=0\qquad ,
\label{determinant}
\end{equation}
while for all extinct 2-mers $\Delta_{ij}>0$. 
%



Now we can put the upper bound on the number of 
surviving 2-mers in the steady state of the system. 
This is accomplished by comparing the total number of constraints 
given by Eqs. \ref{determinant} to the number of 
independent variables. Since matrix $\hat \lambda$ is fixed, 
the only variable parameters in Eqs. \ref{determinant} are the 
left and right end concentrations $l_i(t)$ and $r_i(t)$. 
While naively, the number of such variables is $2Z$, 
Eqs. \ref{determinant} always contain them in combinations 
$l_i \cdot r_{i^*}$. 
Therefore, for the purpose of our counting argument,  
only these $Z$ products should be considered as independent variables. 
The number of constrains (Eqs. \ref{determinant}) that are simultaneously satisfied 
cannot be greater than $Z$. Each of these equations corresponds to either a pair of 
mutually complementary 2-mers or to a single self-complementary 2-mer. 
Denoting the total number of surviving 2-mers as $N$, and 
the number of self-complementary surviving 2-mers as $N_{sc}$, the number of 
equations for surviving 2-mers is given by 
$(N+N_{sc})/2$. Thus the upper bound on the number of surviving 2-mers is given by 
\begin{equation}
N \leq 2Z-N_{sc} \qquad .
\label{2Z_bound}
\end{equation} 
%
%
Note that for large $Z$ the number of surviving 2-mers 
is dramatically lower than $Z^2$ - the total number of possible 
ones.  This explains the entropy reduction observed  numerically (see Fig. \ref{fig1}). 
The parameters of the system were chosen in such a way that initially all $Z^2$ 2-mers 
grow exponentially. 
Since the rate of this early exponential growth depends on 
$\lambda_{ij}$ and $c_i$, it differs from one 2-mer to another. 
This results in a transient behavior where the inhomogeneity of 2-mer concentrations 
is amplified giving rise to an early decrease in entropy (see the dip around $t=1$ in 
Fig. \ref{fig2}a). As concentrations $l_i$ and $r_i$ start to get gradually depleted, the growth 
saturates, giving time for slower-growing 2-mers to catch up with the faster-growing ones 
around $t=2$ (see Fig. \ref{fig2}b). As a consequence, the entropy recovers close to its 
maximal value. After that, a new process starts in which 
2-mers actively compete with each other for remaining left and right ends. When the 
determinant $\Delta_{ij}$ for a particular pair of 2-mers $ij$ and $j*i*$ 
changes its sign to positive, that pair of 2-mers starts to exponentially decay and 
eventually goes extinct. 
This process continues until the number of remaining 2-mers 
$N$ (for which $\Delta_{ij}=0$), falls below the upper bound given by the Eq. \ref{2Z_bound}.
These surviving 2-mers are visible in the heatmap in Fig. \ref{fig2}C.

\subsection*{Graph-theoretical representations}
\begin{figure}
\includegraphics[angle=0,width=0.95\columnwidth]{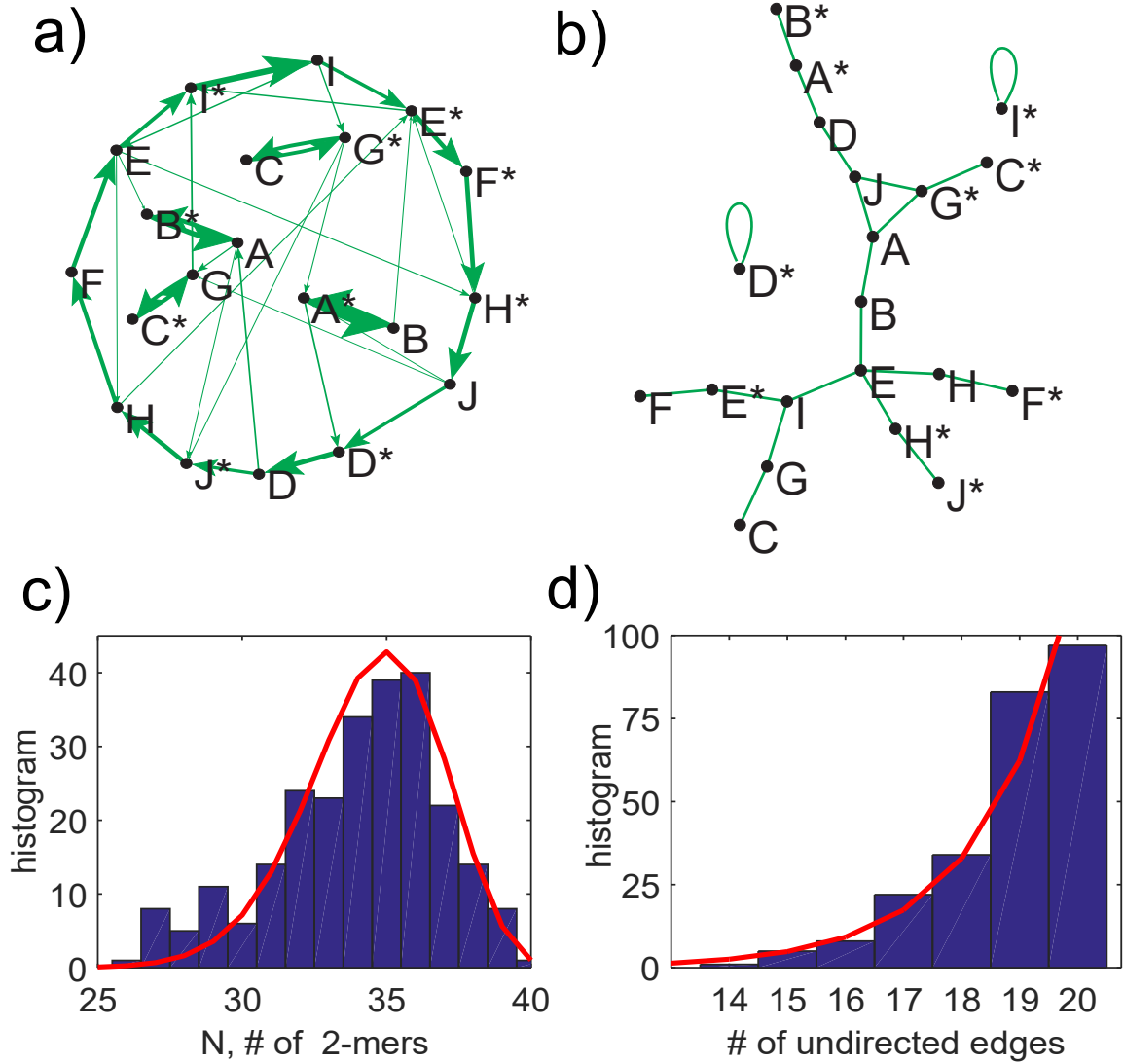}
\caption{
a) The de Bruijn directed graph with $Z=20$ nodes corresponding to monomers and edges corresponding to surviving 2-mers. Thickness of each edge scales with 2-mer's concentration. Polymer sequences are walks on this graph.
b) Undirected graph representation of the system constructed 
as described in the text. Each edge represents two 
mutually complementary 2-mers, for instance, 
$F - E^*$ stands for $F \to E$ and $E^* \to F^*$.  
c) The histogram of the number $N$ of surviving 
2-mers (directed edges in the de Bruijn graph).
d) The histogram of the number of undirected edges $(N+N_{sc})/2$ in 
250 realizations of the model with different $\lambda$'s and $c_i$. 
\label{fig3}
}
\end{figure}

A useful visualization of the emergent state of the system is a so-called 
de Bruijn graph shown in Fig. \ref{fig3}a. It represents each of $Z$ monomer types 
as a vertex, and each of $N$ surviving 2-mers $ij$ as a directed edge connecting 
vertices $i$ and $j$. The weight of every edge is proportional to the steady 
state 2-mer concentration $d_{ij}$. 
A de Bruijn graph is a common representation of heteropolymer ensembles 
such as e.g. DNA sequences of all the chromosomes in an organism. 
It is straightforward to construct it from a known pool of sequences. 
However, the inverse problem of reconstruction of the statistics of a  sequence pool from 
the de Bruijn graph 
is highly non trivial.
Each of the polymers in the pool can be represented as a walk on this graph. 
The simplest case is when the consecutive steps of this walk are 
uncorrelated with each other. This means that the walk is a random Markov process 
with the probability of a step $i \to j$ given by $d_{ij}/c_{i}$, while the 
probability of termination of a polymer at the vertex $i$ given by 
$1-\sum_{j} d_{ij}/c_{i}=l_{i}/c_{i}$. Note that the entropy defined above and 
plotted in Fig. \ref{fig2}a is {\it exactly} the information entropy of a pool of polymer sequences 
generated by such Markov process \cite{andrieux}. Longer-range 
correlations are not captured by the present model, but can in principle could 
emerge due to the effects outlined in the Discussion section,  
leading to a further reduction of the information entropy in the system.

The de Bruijn graph can be complemented by another, more compact,
graphical representation 
specific to our system. Since mutually complementary 
2-mers appear in pairs $ij$ and $j^*i^*$ 
(with the exception of self-complementary 2-mers $ii^*$), 
each such pair can be depicted as a single undirected edge 
connecting vertices $i$ to $j^*$. In this representation each 
edge represents two 2-mers, while each vertex $i$ stands for 
either $i$ or $i^*$ monomer, depending on whether it is the 
first or the second letter within the  
2-mer. For our system this undirected graph (see  \ref{fig2}b)
has a number of remarkable properties derived in SM. 
First, it is a so-called "pseudoforest" \cite{dantzig}: 
each of its individual connected components 
contains no more than one cycle.  This allows us to refine and 
give a topological interpretation to Eq. \ref{2Z_bound}: 
$N=2Z-N_{sc}-2N_{trees}$, where $N_{trees}$ is the number of 
trees (components without cycles) in the pseudoforest. Second, 
only the odd-length cycles (1,3,5, etc.) are allowed in 
this graph. 

Fig \ref{fig3}c shows the distribution of the number of surviving 2-mers (or equivalently of 
directed edges in the de Bruijn graph shown in Fig. \ref{fig3}a) in 250 realizations of the system with 
different values of $\lambda_{ij}$ and $c_i$. 
Fig \ref{fig3}d shows the distribution of the number $(N+N_{sc})/2$ of 
edges in undirected graphs such as one shown in Fig. \ref{fig3}b for these
realizations. As discussed above, the deviation of this last quantity 
down from $Z$ is equal to the number of trees in the pseudoforest. 
As shown in Fig. \ref{fig3}d, it follows approximately 
exponential distribution with the average around 1.6 
(the red line in Fig. \ref{fig3}c). 
At the same time, the distribution of the 
number of surviving 2-mers ($N$) has a peak around $36<2Z=40$. The 
quantity $2Z-N$ is always positive and 
approximately follows a Poisson distribution 
with the average of $5.5$ (red line in Fig. \ref{fig3}c).  


\subsection*{Variability of the surviving set of 2-mers}
The set of surviving 2-mers and their concentrations 
depend on a number of parameters: ligation rates $\lambda_{ij}$, 
total monomer concentrations $c_i$, and, possibly, seed concentrations of 
individual 2-mers. 

We analyzed the sensitivity of the steady state 
of our system with respect to all of these parameters one-by-one. First we fixed both $\lambda_{ij}$,
and $c_i$ and analyzed the final 2-mer concentrations for a large number of random realizations of 
$Z^2$ small (but positive) seed concentrations.
We found the final state to be completely reproducible as long as all seed concentrations are 
non-zero. Note that due to autocatalytic nature of 2-mer dynamics, Eq. \ref{eq_2-mer_dynamics}, 
a pair of complementary 2-mers with zero seed concentrations would never emerge on 
their own. 

\begin{figure}
\includegraphics[angle=0,width=0.95\columnwidth]{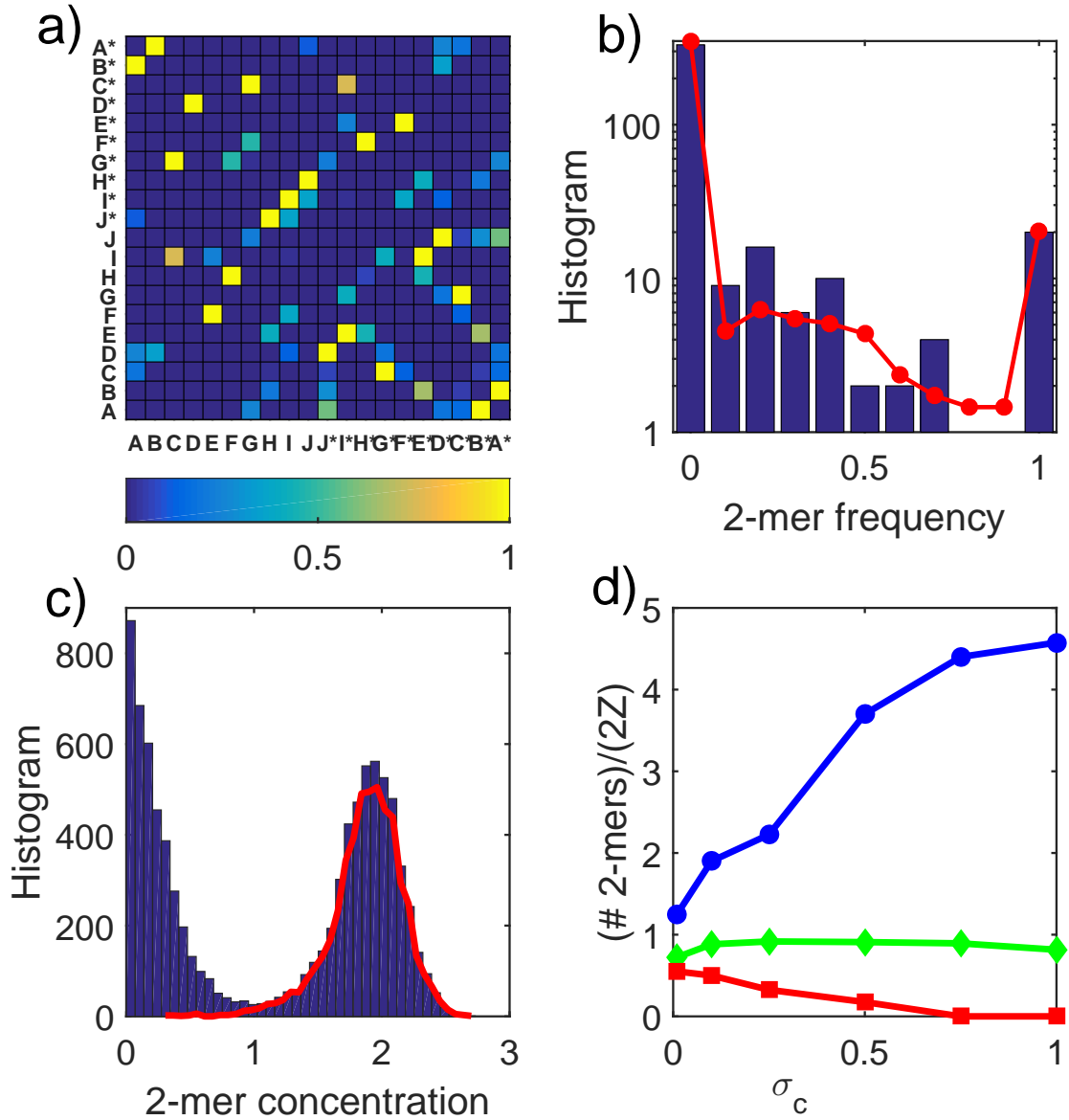}
\caption{
a) The heatmap of survival frequency of 2-mer in the steady state of the system with 
fixed $\lambda_{ij}$  (same as in Figs. \ref{fig2}-\ref{fig3}) and 100 
different realizations of $c_i$. Note the similarity to Fig. \ref{fig2}c.
b) The histogram of these survival frequencies (blue bars) as well as 
its average (red line) over 
10 different realizations 
of $\lambda_{ij}$. 
c) The histogram of 2-mer concentrations in 250 realizations of 
$c_i$ for fixed $\lambda_{ij}$. The red line shows the contribution from 
20 "universal survivors" - 2-mers present across all realizations.
d) The number of universal survivors (red line), 
the average number of survivors in a single realization (green line) 
and the number of 2-mers present in at least one among hundred 
realizations of $c_i$ (blue) each normalized by $2Z$. The x-axis 
is the width $\sigma_c$ of log-normal distribution of monomer concentrations.  
\label{fig4}
%
%
%
}
\end{figure}

Next, we fixed the ligation rates $\lambda_{ij}$ to their values used to 
construct the heatmaps shown in Fig. \ref{fig2}
and networks in Fig. \ref{fig3}. We then simulated 100 realizations of the system with 
$c_i$ pulled from a log-normal distribution
$P(c_i) \sim \exp(-[\log(c_i/c)]^2/\sigma_c)/c_i$ with $c=3$ and $\sigma_c=0.1$. 
Fig. \ref{fig4}a shows the heatmap of the fraction of realizations of $c_i$ in which each individual 
2-mer survives in the steady state. In Fig. \ref{fig4}b we present the same results in the form of the 
histogram (blue bars). The majority of 2-mers (324 out of 400 visible as the leftmost bar in Fig. \ref{fig4}b) 
got extinct in all of 100 realizations. 
While the number of surviving 2-mers in each realization never exceeded $2Z=40$ (see Fig. 
\ref{fig2}c), the overall number of 2-mers that survived in at least one realization of $c_i$ 
was substantially larger: 76. 
Out of them, 20 "universal
survivors" were present in all 100 realizations. 
Furthermore, as can be 
seen by comparing heatmaps in Figs \ref{fig2}d and Fig. \ref{fig4}a, these 20 2-mers
typically have high steady state concentrations $d_{ij}$. To further investigate 
the correlation between 2-mer concentration and its survivability, 
we analyzed a larger set of 250 realizations of the system with the same 
fixed ligation matrix but variable monomer concentrations $c_i$. 
The resulting distribution of 2-mer concentrations shown in Fig. \ref{fig4}c is clearly bimodal. 
The bimodality is also apparent from an example of de Bruijn graph shown in Fig. \ref{fig3}a, 
where approximately half of all links (thick lines) 
correspond to more abundant 2-mers, while the other half (thin lines) - to 2-mers 
present at low concentrations.  The high-concentration peak of the distribution 
in Fig. \ref{fig4}c 
is dominated by the contribution from 
20 universal survivors shown as red line. Note that despite 
an increased number of realizations, the set and number 
of universal survivors stayed the same as in Fig. \ref{fig4}a-b. 

We further investigated the variability of the set of surviving 2-mers as a function of 
width $\sigma_c$ of the log-normal distribution of monomer concentrations. As shown 
in Fig. \ref{fig4}d, the number of universal survivors (red line) systematically decreases with $\sigma_c$,  
ultimately reaching 0 for $\sigma_c \geq 0.75$. 
Consistent with this trend, the bimodality of the 
concentration distribution in Fig. \ref{fig4}c 
also disappears for larger valu$\sigma_c$.   
Note that all numbers of 2-mers shown in  Fig. \ref{fig4}d were 
normalized by $2Z$, i.e. the upper bound of the number of surviving 2-mers in each realization. 
The average number of survivors in a single realization (green line)
does not have a notable dependence on $\sigma_c$. The blue line in 
Fig. \ref{fig4}d shows 
the number of 2-mers (normalized by $2Z$) that survived
at least once among 100 realizations of $c_i$.
Note this curve grows significantly with $\sigma_c$ reaching 
the value as high as 5. This corresponds to half of all 
$Z^2=400$ possible 2-mers having a chance to survive at 
in least one of these realizations. 
\section*{Discussion}
The major conclusion following from this study is that our model
system of mutually catalyzing heteropolymers has a natural 
tendency towards spontaneous reduction of the information entropy. 
This represents an effective {\it reversal} of the second law of 
thermodynamics in this class of systems. While {\it "violations"}
of the second law are indeed expected in externally driven 
non-equilibrium systems, the observed "reversal" has much greater implications.
Both living organisms and other self-organized systems 
such as human culture, economics, and technology, 
are characterized by an ever increasing complexity, indicating the ongoing 
reduction in the information entropy. 

The thermodynamic entropy of a system of heteropolymers 
is composed of two distinct parts \cite{andrieux}: 
(i) the translational and configurational entropy of 
polymer chains; 
and (ii) the information entropy associated with sequence statistics. 
Our current model hints at a hierarchical scenario of 
entropy reduction in populations of heteropolymers. 
First, the translational entropy is reduced due to 
template-based polymerization as studied in our previous work 
\cite{tkachenko2015} within Random Sequence Approximation (RSA). 
Then RSA breaks 
down at the level of 2-mers due to their competition 
with each other for a limited 
resource of monomers. Such symmetry breaking in the sequence 
space results in a dramatic reduction in the information entropy (ii).  
%
%
%
At this point, sequences of the entire pool of chains are 
generated as Markovian random walks on the de Bruijn graph 
(Fig. \ref{fig3}a). One can imagine a further 
reduction of the information entropy due to the emergence of correlations 
between consecutive steps of this walk. 

There are multiple 
physical scenarios outside the scope of our current model 
that would lead to such longer-range 
correlations in the sequence space. 
They include e.g. a dependence of association rates in 
Eq. \ref{eq_2-mer_dynamics} on lengths of the three chains
involved in the process of template-assisted ligation. 
Another intriguing scenario is 
a spontaneous emergence of chains with weak catalytic activity above 
and beyond their role as templates for ligation. For instance, some sequences 
may facilitate breakage or conversely promote ligation reactions, either among some 
specific sequences or universally. Such sequences would provide a missing link 
between the prebiotic soup considered here and the emergence of the  first ribozymes in 
the RNA world scenario. 

A common pattern in functional RNA-based structures such as ribozymes and ribosomes is 
the presence of hairpins and loops. 
Note, however, that mass-action term in the 
Eq. \ref{eq_2-mer_dynamics} assumes that the template and the two substrates 
belong to three different chains thus ignoring these higher order structures. 
A proper model description taking them into account is a natural next step 
in development of our approach. 
An important future question to address is whether 
our system would naturally evolve towards or away from loops and hairpins.
One might expect that a hairpin composed from 
several pairs of mutually complementary 2-mers would 
have a self-healing property: any broken bond would be quickly repaired 
since the template and one of the substrates belong to the same chain
and hence  are more likely to be hybridized. 

%
An interesting feature of the sequence statistics emerging in our model 
is that the entropy is not reduced to its absolute minimum that would 
correspond to a unique "master sequence". In the de Bruijn representation such 
master sequence would look e.g. like a single cycle (or several unconnected cycles) 
in which for every monomer there is unique right neighbor following it on every 
chain. In this case the walk on the de Bruijn graph would be completely deterministic.
In contrast, in our model each monomer typically has two possible right neighbors
in de Bruijn graph. In the limit of relatively small 
variations $\sigma_c=0.1$ used to construct the Fig. \ref{fig3}a, 
stronger links corresponding to "universal survivors", produce 
$c_i$-independent backbone of the graph akin to the master sequence. 
%
Conversely, weaker links allow for infrequent hopping between different parts of 
the graph leading to deviations from this master sequence. 
The opposite limit of large variations in monomer concentrations 
is especially promising from the point of view of further 
evolution.
In that limit, there is  no dominant master sequence. 
This dramatically expands the explored region in the sequence space: 
now, for every step of the Markovian walk there are on average two 
comparable probabilities for selecting the next monomer. 
As a result, the number of possible sequences of length $L$ 
in our model, $\sim 2^L$, remains exponentially large, yet 
dramatically reduced compared  
to its random sequence limit, $Z^L$.
Furthermore, in the limit of large $\sigma_c$,
different realizations of $c_i$ give rise to significantly different sets 
of surviving 2-mers. Note that, unlike the ligation rates $\lambda_{ij}$, 
monomer concentrations $c_i$ (or equivalently their influxes 
$\phi_i$) could vary significantly from one spatial 
location to another (See \cite{Hud2012} for a study 
of spatial inhomogeneity in the prebiotic context). 
This allows for an effective exploration of 
various regions (of size $\sim 2^L$ each) of the global 
sequence space, rather than converging to the same
subset of sequences over and over. 


Our model describes a simple yet general mechanism for 
spontaneous entropy reduction in systems capable of 
template-assisted ligation. There are multiple possible 
experimental realizations of such systems based on either 
traditional DNA/RNA biochemistry or artificial micro/nano
structures. The most direct implementation of our model would 
be a system composed of $Z$ words made of 
a string of nucleotides bound together by strong (e.g. DNA-type) bonds. 
They have to be designed to form $Z/2$ mutually complementary pairs that are orthogonal 
to each other, i.e. words from different pairs have no 
long overlaps. These words would play the role of composite monomers 
in our model that could be subsequently connected to each other with weaker, 
breakable (e.g. RNA-type) bonds.  
%
As discussed earlier, our model is directly applicable to the scenario 
in which all bonds are unbreakable, while the whole system is uniformly
diluted and fresh (unbound) monomers are supplied at a constant rate.  
This greatly expands possibilities for its experimental implementation.
Similarly, our dynamics can be 
implemented e.g. using the DNA origami nanoblocks introduced 
in Ref. \cite{Chaikin}. It should be emphasized, that in order to achieve 
the behavior described by our model, the experiments need to be conducted 
well below the saturation regime, i. e. the night phase should be 
shorter than a typical association time for hybridization.


Yet another interpretation of our model does not involve any long polymers at all.
In this case 2-mers are represented by physical dimers made of only two 
monomeric units incapable of forming longer chains. Our model predicts that, even 
in this simple system, the compositional entropy would drop because of the extinction 
of most of the dimers leaving only $N\leq 2Z$ survivors.
On the one hand, this further extends possibilities for experimental 
implementation. For instance, one could construct the DNA-based system described above but 
limited to no more than  2-word chains. This would greatly reduce the complexity of the 
screening process. 

On the other hand, this dimer interpretation has 
an intriguing connection to the Kauffman model of autocatalytic chemical reaction 
networks \cite{Kauffman}. In the Kauffman case, some of the molecules types in 
the pool are capable of catalyzing the synthesis of others from "raw materials" 
(abundant small metabolites) 
ultimately resulting in the emergence of metabolic autocatalytic cycles 
in large systems. 
A recent model of such chemical reactions \cite{EnglandPNAS2017} 
shows that the system self-organizes to a state finely tuned to the external 
driving force. This can be interpreted as maximization of the rate of 
negentropy adsorption from the environment. 
In our case dimers correspond 
to mutually catalytic entities while monomers represent raw materials. While 
in the current implementation of our model, the catalytic cycles could only involve 
two mutually complementary dimers, it is straightforward to generalize the model 
to allow the mutual catalysis of  any pair of dimers. We expect our findings about the reduction of entropy to be fully 
transferable to that case. Thus, our model has a potential of bridging 
the gap between two traditionally competitive visions of the Origin 
of Life: "information first" and "metabolism first".

\subsection*{Acknowledgements}
This research used resources of the Center for Functional Nanomaterials, 
which is a U.S. DOE Office of Science User Facility, at Brookhaven National Laboratory under Contract No. DE-SC0012704.


\newpage
\clearpage
\pagenumbering{arabic}
\setcounter{page}{1}
\section*{Supplementary Materials}
\subsection*{Equivalence between 
bond breakage and dilution}
Here we demonstrate that the set of equations 
(\ref{eq_2-mer_dynamics}-\ref{eq_li} ) also describes 
the system in which intra-polymer bonds are unbreakable, while concentrations 
of all molecules are diluted at rate $\beta$ and fresh monomers 
of type $i$ are supplied at rate $\phi_i=\beta \cdot c_i$. 
Indeed, the dilution adds terms $-\beta \cdot d_{ij}$
to the r.h.s. of equations \ref{eq_2-mer_dynamics}. The dynamics of 
individual monomer concentrations (both bound and unbound), $c_i$, 
are given by equations $\dot{c}_{i}(t)=
\phi_i-\beta c_i(t)$.
After a brief transient regime, all monomer concentration 
$c_i(t)$ reach a steady state value $\phi_i/\beta=c_i$ so that   
Eqs. \ref{eq_li} become automatically satisfied. 
In this way the problem with dilution is exactly mapped 
onto the problem where bond breakages are allowed. 

\subsection*{Topological analysis of the undirected graph}
Here we discuss the undirected graph representation 
of the pool of heteropolymers. 
Since our system always contains pairs of mutually complementary 2-mers 
$ij$ and $j^*i^*$ (with exception of self-complementary 2-mers $ii^*$), one can represent 
this pair with a single undirected edge connecting $i$ to $j^*$. These edges form an 
undirected graph shown in Fig. \ref{fig3}b. 
Note that due to these rules
an edge connecting vertices $i$ and $k$ in this graph represents a pair of 
2-mers $ik^*$ and $ki^*$ shown as two edges in Fig. \ref{fig3}a 
or two matrix elements in Fig. \ref{fig2}c. For simplicity in Fig. \ref{fig3}b 
we did not assign weights to these symmetric edges. 
Each edge of this graph corresponds to exactly one equation in the set of 
Eqs. \ref{determinant}. Hence, the number of undirected edges is equal to   
$(N+N_{sc})/2$ and according to Eq. \ref{2Z_bound} it cannot exceed $Z$.
On the other hand, based on network topology, this number can be expressed as 
$Z-N_{comp}+N_{cycles}$. 
Here $N_{comp}$ is the number of connected components of the graph, while 
$N_{cycles}$ - the number of independent cycles defined as the minimal number of 
edges one needs to cut to remove all cycles.  The inequality  (\ref{2Z_bound})
means that the number of independent cycles cannot be larger than the number of 
components. Furthermore,  this inequality must be also satisfied for each of the 
individual connected  components of the graph because the number of 
equations (edges) cannot exceed the number of independent variables 
(the number of vertices in this components).  In other words, 
{\it each of the components may 
contain no more than one cycle}.  Graphs with this property are known as  
"pseudoforests" \cite{dantzig}. For unicyclic components (i.e. 
those that include exactly one cycle), the numbers of edges and vertices are equal to 
each other, while for each of the tree (cycle-free) components, 
the difference between these two numbers is equal to $1$. This gives a 
topological interpretation to the number of surviving 2-mers $N$ in 
terms of the number of tree-like components of the undirected graph, $N_{trees}$: 
\begin{equation}
N=2Z-N_{sc}-2N_{trees} \quad ,
\end{equation}
which automatically leads to inequality  (\ref{2Z_bound}).

One can also demonstrate that only the cycles of odd lengths (1,3,5, etc.) 
are allowed in our system. 
Indeed, for a  hypothetical even-length cycle 
$i_1-i_2^*-i_3-i_4^*-...i_{n-1}-i_n^*-i_1$, 
one can construct a combination of Eqs. \ref{determinant} of the following form: 
\begin{equation}
\frac{\Lambda_{i_1i_2}\Lambda_{i_3i_4}...\Lambda_{i_{n-1}i_{n}}}
{\Lambda_{i_2i_3}...\Lambda_{i_{n-2}i_{n-1}}\Lambda_{i_ni_1}}=1
\end{equation}
Here $\Lambda_{ij}=1-\Delta_{ij}=\lambda_{ij}  \cdot \lambda_{j*i*} 
\cdot l_i \cdot r_j  \cdot  r_{i*}  \cdot l_{j*}$
All the variables 
$l_i$ and $r_i$ at the left-hand-side of this equation cancel, 
making it an  invariant that depends only on ligation 
rates.  Therefore this equation cannot be satisfied for 
a generic matrix $\lambda_{ij}$, which rules out the existence of 
even cycles in most of the cases.


\begin{thebibliography}{10}
\bibitem{disclaimer} Strictly speaking, due to the exchange of energy and material with the environment, 
the appropriate measure of the reduction of thermodynamic entropy by living systems is the
local accumulation of Gibbs free energy. This was pointed out by Schr\"{o}dinger in Ref. [3].
\end{thebibliography}
\end{document}